# A NEW MODEL FOR THE ORGANIZATIONAL KNOWLEDGE LIFE CYCLE

Luigi Lella, Ignazio Licata•

ISEM, *Institute for Scientific Methodology, Palermo.Italy*

**ABSTRACT**

Actual organizations, in particular the ones which operate in evolving and distributed environments, need advanced frameworks for the management of the knowledge life cycle. These systems have to be based on the social relations which constitute the pattern of collaboration ties of the organization. We demonstrate here, with the aid of a model taken from the theory of graphs, that it is possible to provide the conditions for an effective knowledge management. A right way could be to involve the actors with the highest betweeness centrality in the generation of discussion groups. This solution allows the externalization of tacit knowledge, the preservation of knowledge and the raise of innovation processes.

## 1. THE KNOWLEDGE LIFE CYCLE

Nowadays every organization must be able to learn quickly and continually from the environment where it operates [Nonaka, 1994]. The new knowledge comes from the experiences of the individuals operating within the organization, and it is constructed through their social and collaborative interactions [Nonaka and Takeuchi, 1995; Nonaka, Toyama and Konno, 2000]. For this reason the technology should focus on the problem of finding innovative solutions to improve the cooperation among individuals and the awareness of the knowledge and the skills reached by each of them [Stenmark, 2003].

As noticed by McElroy [McElroy, 2003] the knowledge management systems (KMS) of the first generation have been focussed principally on the processes of knowledge diffusion and integration. This means they were based on the assumption that valuable knowledge was already present within the organization. Therefore the main purpose of a KMS was to provide the right information to the right people and to codify all the explicit and tacit knowledge embodied in organizational processes and in the beliefs of the individuals.

But the purpose of a KMS should be also the production of new knowledge, not only the integration of the existing organizational knowledge. So, as stated by McElroy, the KMS of the second generation have all to deal with the problem of the creation of knowledge, favouring the detection of problems and needs and the finding of solutions.

This innovation process proceeds in a form that is called by McElroy "Knowledge Life Cycle" or KLC. The KLC is not just a model, but a "framework for placing models in context" using the exact definition of McElroy. So a complex of different competing ways and views of how knowledge can be produced and integrated. The way this framework works is influenced by the following assumptions.

First of all the learning foundation is the experience of gaps in everyday activities. The detection of these gaps, which are the lack of the needed knowledge to carry out the activities in the right way and in the shortest time, represents a sort of emergence of problems.

---

• corresponding author: Ignazio.licata@ejtp.info

The detection of gaps is just the first step toward the formulation of the problems which are defined by McElroy "knowledge claims", which comprise an analysis, an elaboration of the problems to be solved. Knowledge claims could be conjectures, assertions, reports, guidelines or entire theories on the right processes to follow in order to fill the detected gaps.

The formulation of a knowledge claim can involve more individuals leading to the generation of groups. These communities, in a formal or informal way, share ideas submitting them to a sort of peer review. This process is requested to validate the emergent innovative ideas. Such process of knowledge claim formulation and evaluation is considered by McElroy a process of knowledge production.

Not all the knowledge claims survive within the organization finding the interest and the approvation of the other individuals. The ones that do not succeed in the evaluation process could be "undecided knowledge claims" or "falsified knowledge claims". The reports which certify the failure of the knowledge claims are called "meta-claims" i.e. claims about claims.

The knowledge claims which pass the validation are instead integrated into the activities of a wider group of people. The integrated knowledge can take the form of mentally held knowledge by individuals or groups or explicit artifacts like documents and files. The first type of container can be considered a special kind of tacit knowledge, but only the explicit forms of knowledge are considered "knowledge claims" by McElroy.

The following phase is the knowledge use which regards the business processing, not the knowledge processing, even if new problems, and so other knowledge claims, can arise even in this last phase.

Summing up the processes of Knowledge Production, Knowledge Integration and Business Processing have not to be conceived as isolated, but they interact each other in a complex manner. And their complexity degree has to be realized in order to support the organizational processes of innovation. Means that the capture, the coding and the deploying of knowledge alone are not sufficient to guarantee the creation of innovation. These efforts are merely examples of information management or information processing, not knowledge management.

The main intuition of McElroy is that a KMS has to guarantee strategies and environments where knowledge can also be valuated producing knowlegde claims and meta-claims. Only an evaluative and critical process can integrate and coordinate the different phases of knowledge management.

## 2.A NEW FRAMEWORK TO SUPPORT KLC

According to the definition of McElroy, in order to support the knowledge life cycle the KMS has to provide and promote knowledge sharing spaces where individuals can discuss certain problems, conjectures and theories.

The system we are going to present tries to achieve this goal in two steps. First the social network of the entire organization is analyzed to detect the points where knowledge and information pricipally flow. Once these individuals have been detected they are prompted by the system to create a community to discuss a given knowledge claim of common interest.

This group can take the form of a community of practice [Hildreth and Kimble, 2000; Wenger, McDermott and Snyder, 2002; Saint-Onge and Wallace, 2003] where individuals meet each other in face to face encounters or it can take the form of a network of practice [Hildreth and Kimble, 2004] where individuals take on the debate in virtual environments as forums, blogs [Jensen 2003] or a wiki [Ebersbach, Glaser and Heigl, 2005].

The encounter has to produce a document, for example a report, a guideline, a directive, which resumes the ideas, the problems and the solutions which have emerged in the debate. This document has to be structured as an hypertext, meaning that it has to contain also references to other documents and reports. The network of documents can be considered as a network of ideas which have been externalized by a single author or a community of authors as the discussion reports.

It has to be stressed that the present work doesn't want to cope with the problem of the definition of an opinion formation model [Di Mare and Latora, 2006; Bordogna and Albano, 2007]. It wants only to present a preliminary study on the effects of the introduction of a new knowledge management platform on the evolution of networks of ideas and knowledge claims within an organization.

Figure 1 shows the two different dimensions taken into consideration by the system, which are the organizational social network and the network of ideas which emerge from the knowledge production spaces provided by the system and maintained by the individuals which intercept the majority of knowledge and information flows.

The emerging network of ideas can be conceived as a complex system which constantly evolves in time. This implies that the structure of the network continuously changes through the addition or the removal of nodes and links. In this kind of networks the survival of nodes seems to depend on some quality of the nodes, for example the quality or the perceived interest or utility of the exposed idea or knowledge claim. Thanks to the innovative value of the claimed ideas it can happen that some research papers in a short timeframe acquire a very large number of citations, much more than other contemporary or older publications.

As stated by Bianconi and Barabasi [Bianconi and Barabasi, 2000] this example suggests that the nodes of a network of ideas have a different ability (fitness) to compete for links. The success of the idea depends also on its popularity and its foundations which are represented respectively by the number of other documents which reference the externalized idea (for example the in-bound links of an electronic hypertext) and the number of document referenced by the externalized idea (for example the out-bound links of an electronic hypertext).

A good model which considers all these factors is the one presented by Bianconi and Barabasi [Bianconi and Barabasi, 2000]. The process starts with a net consisting of N disconnected nodes. At every step t=1..N each node establishes a link with other m units. If j is the selected unit, the probability that this node establishes a link with the unit i is:

$$P_i = \frac{U_i k_i}{\sum_j U_j k_j} \qquad (1)$$

where $k_i$ is the degree of the unit i, i.e. the number of links established by it, while $U_i$ is the fitness value associated to the node.

We want to define such network principally by connecting the knowledge claims which have been externalized in debates and encounters promoted by the individuals which control the flows of information and knowledge. This choice is due to the fact that these individuals have a broader and more generalized vision of the problems and needs of the organization. So they are more capable to suggest knowledge claims which inspire the interest of a wide organizational community.

The betweeness centrality has been considered in literature [Marsden, 2002; Alony, Whymark and Jones 2007] as a way to find the most valuable nodes within a social network. It can be said that a node with an high betweenes centrality plays a "broker" role in the network, i.e. it has a great influence over what flows (and does not) in the network. These nodes play indeed an important role but at the same time constitute a failure point of the network. That is because without their presence some subgroups of individuals within the organization could be cut off from information and knowledge. The betweeness centrality $b_i$ of a node i belonging to a social network is obtained as:

$$b_i = \sum_{j,w} \frac{g_{jiw}}{g_{jw}} \qquad (2)$$

Where $g_{jw}$ is the number of shortest paths from node j to node w (j,w ≠ i) and $g_{jiw}$ is the number of shortest paths from node j to node w passing through node i. The betweeness centrality is indeed the best way to select people which can start and promote discussion among the other individuals within the organization.

The main purpose of our system is twofold. First of all our platform has to promote environments where people can share their ideas on topics of common interest. To achieve the best results the system detects the most important people in the knowledge life cycle, that is those with the highest betweeness centrality. These people have to suggest (maybe with the aid and the prompt of the system) a knowledge claim which can regard the largest audience. All the suggestions, problems and solutions which are emerged from the community that grows around the promoters of the discussion are grouped, organized and externalized. In this way a certain amount of tacit knowledge, that is knowledge principally held in the minds of individuals or embedded in processes [Polanyi M., 1967; Nonaka I., 1994; Nonaka et al., 1998; Hildreth and Kimble, 2000; Bhatt, 2001; Bosua and Scheepers, 2002], can be exteriorized in an explicit form like a report or a guideline. Thanks to such sharing environments new ideas can arise promoting the creation of knowledge and innovation.

At the same time the system allows to achieve another important outcome that is the preservation of knowledge. During the encounters the participants can get in touch with people never seen before or people whom they have never collaborate with. In this way the knowledge can survive even without the promoter of the discussion. For this reason it has to be considered that the fitness function, that is the betweeness centrality, is not constant but changes over time. The participation to a common activity as a discussion forum or the updating of the content of a collaborative work environment as a wiki or a community blog can be conceived as a form of communication or social relation. So in our model the fitness function we choose for a given externalized idea takes the following form:

$$b_i(t) = U_i(t) = \begin{cases} 0 & t < t_i \\ \sum_{j,w} \frac{g(t)_{jiw}}{g(t)_{jw}} & t \geq t_i \end{cases} \quad (3)$$

where $t_i$ is the instant at which the discussion group is constituted. It implies that the betweeness centrality of each node of the social network may vary over time thanks to the effect of the collaborative process of knowledge creation. We assume that $b_i(t)$ is a decreasing function for $t>t_i$, considering that the creation of a discussion group involves a rewiring process in the social network localized around the node i which promotes the discussion. This factor has important effects on the evolution of the network of opinions and ideas as we will demonstrate hereafter.

In another work Bianconi and Barabasi [Bianconi and Barabasi, 2001] compared their model to the evolution of a Bose gas, assigning an energy $\varepsilon_i$ to each node determined by its fitness $U_i$ and a parameter $\beta$ acting as an inverse temperature ($\beta=1/T$):

$$\varepsilon_i = -\frac{1}{\beta}\log U_i \quad (4)$$

According to this mapping, a link between two nodes i and j with different fitnesses $U_i$ and $U_j$ corresponds to two different noninteracting particles on the energy levels $\varepsilon i$ and $\varepsilon j$. The addiction of a new node to the network corresponds to the insertion of a new energy level $\varepsilon_i$ and 2m particles to the gas. In particular m particles, corresponding to the m out-bound links of node i, distribute themselves on the level $\varepsilon_i$ while the other m particles are deposited on other energy levels corresponding to the in-bound links coming from node i. The probability that a particle is settled on a level i is given by (1), and deposited particles are not allowed to jump to other energy levels.

Each node added at time $t_i$ and corresponding to an energy level $\varepsilon_i$ is so characterized by an occupation number $k_i(\varepsilon_i,t,t_i)$ representing the number of links (particles) that the node establishes at time t. Bianconi and Barabasi [Bianconi and Barabasi, 2001] made the assumption that each node increases its connectivity following the power law:

$$k_i(\varepsilon_i,t,t_i) = m\left(\frac{t}{t_i}\right)^{f(\varepsilon_i)} \quad (5)$$

By the introduction of a chemical potential $\mu_i$ they also demonstrated that the dynamic exponent $f(\varepsilon_i)$ takes the following form:

$$f(\varepsilon_i) = e^{-\beta(\varepsilon_i-\mu_i)} \quad (6)$$

This mapping has lead to the prediction of the existence of three different phases in the evolution of their network model.

When all the nodes have the same fitness (6) predicts that $f(\varepsilon_i)=1/2$ and according to (5) the occupation number, which corresponds to the connectivity of node i, increase as $(t/t_i)^{1/2}$. This means that old nodes having smaller $t_i$ have larger $k_i$ and the model reduces to the scale free model [Albert and Barabasi, 2001]. In our case this result indicates that new ideas tend to originate from the most popular ones, which establish more connections with the others, and old ideas have more chances to become more popular and survive than the others. Clearly this phase, that has been called by Bianconi and Barabasi "first-mover-wins" (FMW), doesn't correspond to a real network of opinions where the value of the idea influences more its success than its age.

In systems where nodes have different fitnesses the fittest nodes acquire links at an higher rate even if they have been introduced at a later time with respect to the others. This phase is called by Bianconi and Barabasi "fit-get-rich" (FGR). In our model the value of an externalized idea is approximated by the fitness (3), which can be considered the extent by which the promoter of the knowledge claim is capable to interest the members of the community which discuss the knowledge claim.

But in the two first phases there is no clear winner as the fittest node's share of all links decreases to zero in the thermodynamic limit, leading to the emergence of a hierarchy of few large hubs surrounded by many less connected nodes. In the "first-mover-wins" phase the relative connectivity of the oldest node follows the law:

$$\frac{k_{max}(t)}{mt} \approx t^{(1/2)-1} = t^{-1/2} \to 0 \quad (7)$$

In the "fit-get-rich" phase the relative connectivity of the fittest node decreases as:

$$\frac{k(\varepsilon_{min},t)}{mt} \approx t^{f(\varepsilon_{min})-1} \to 0 \quad (8)$$

considering that $f(\varepsilon_{min}) < 1$. Bianconi and Barabasi demonstrated that below a given $T_{BA}=1/\beta_{BA}$ the fittest node maintains a finite fraction of the total number of connections during the growth of the network. This particular phase has been compared by Bianconi and Barabasi to the Bose-Einstein (BE) condensation [Huang, 1987].

In a network of knowledge claims this phase has to be avoided because it means that a given idea prevails over the other ones, limiting the process of innovation of the organization. As stressed by Bianconi and Barabasi real networks have a T independent fitness distribution meaning that their status (BE or FGR) is independent of T. Luckily our KMS model tends to level the fitnesses of the nodes. The rewiring of social ties around the individuals with the greatest betweeness centralities leads to the appearance of new individuals with the highest betweeness centralities. In this way the BE condensation can be avoided. This assert cannot be mathematically demonstrated as the fitness function depends on an unspecified number of variables, but figure 1 and figure 2 can show the way by which the network of externalized ideas evolves in time.

At t=0 we have three individuals A, B and C with the highest betweeness centralities which are invited by the system to promote a space of discussion reporting all the arisen knowledge claims in the documents A, B and C. Some people within the discussion groups A and B notices a correlation among the themes treated by A and B and a reference is generated among the corresponding reports. A correlation is detected between the reports B and C and another reference is added. The spaces of discussion are open to every interested participant and alerting measures could be adopted in order to spread the invitations over the entire organization. In this way there can arise large communities which do not include only the strongest ties of the discussion promoter.

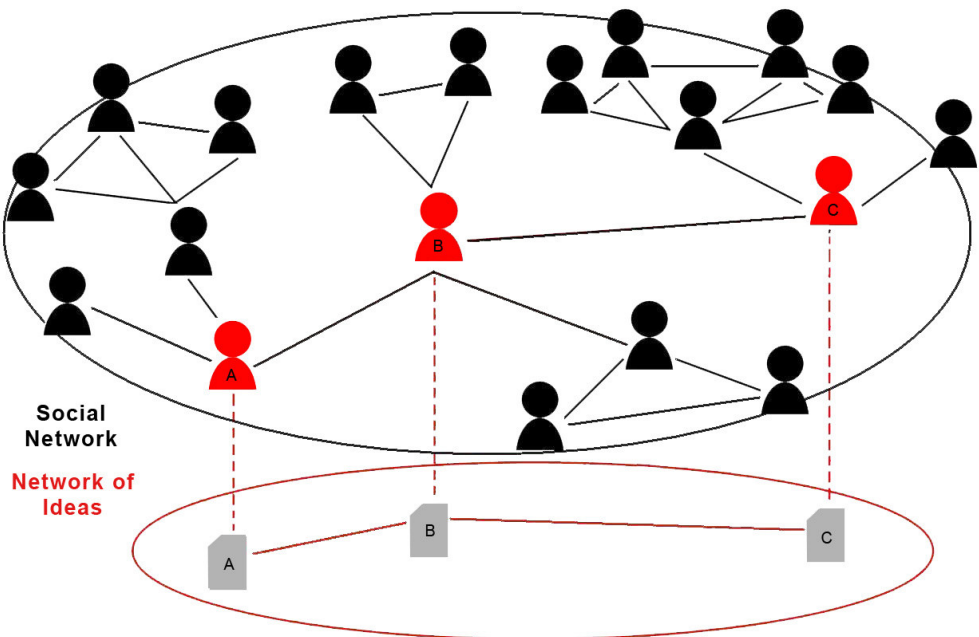

Figure 1 – Network of externalized ideas at t=0

At t=1 the pattern of ties within the social network is changed. The betweeness centrality of A, B and C has decreased and the individuals D and E, having the highest betweeness centrality, are invited to promote other two discussion spaces. Individuals A, B and C continue to attend to the discussions of their groups, but the interest on their knowledge claims has vanished. This is attested by the decrease of the betweeness centrality of the promoters A, B and C. Thanks to the high fitness the node E within the network of externalized ideas can establish the same number of connections as node B. Probably the individual E promotes a meta-claim on the knowledge claim sustained by the individual B. This justifies the presence of the connection among reports B and E.
But the ever changing values of the business centralities of the promoters guarantees that no externalized knowledge claim will prevail over the other ones.

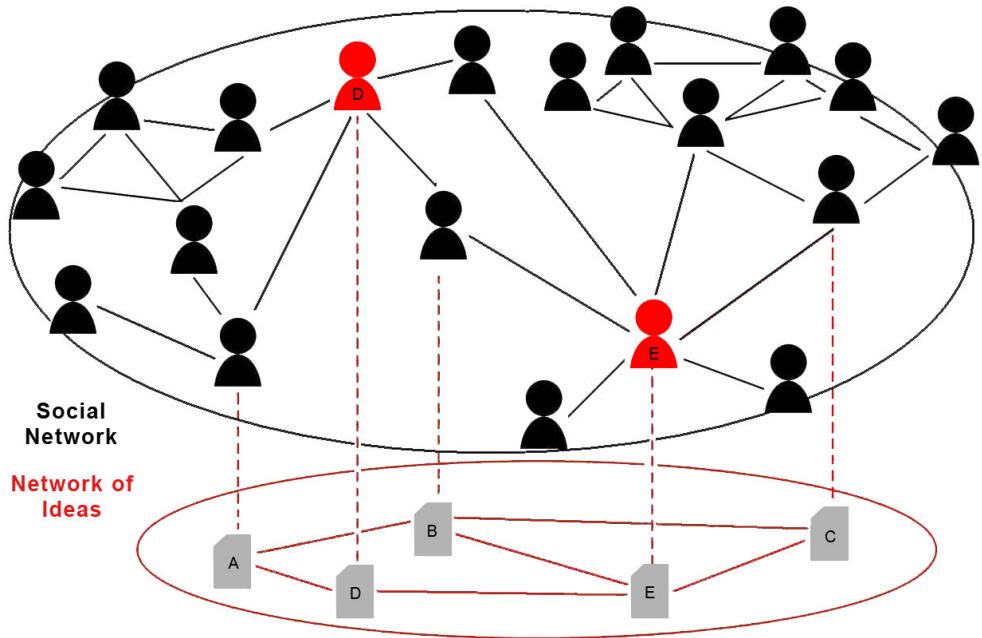

Figure 2 – Evolution of the network of externalized ideas at t=1

## 3. CONCLUSION AND FUTURE WORK

A KMS needs techniques and strategies to support the entire knowledge life cycle. This process has to lead to the formulation of knowledge claims and meta-claims, which are produced by problem analysis and problem validation processes. These activities cannot be scheduled and structured beforehand in a top-down fashion by the management, but they have to arise in an emergent manner, considering the knowledge gaps encountered by the agents in their activities and involving the right people which can effectively judge and deal with the arisen knowledge claims.

A possible way could be to monitor the evolution of the social network which characterizes the organization in order to detect the individuals with the highest betweeness centralities and prompt them to detect problems. These individuals intercept the majority of information and knowledge flows and therefore they are the rightest people to suggest knowledge claims, submitting them to the pair review of a large involved and interested community. In other words these actors are invited by the system to produce knowledge. The discussion environments promoted and sustained by these individuals drives a social relation complexity turned to the generation of meta-claims or other knowledge claims.

In this work we presented a knowledge management framework that is designed to externalize tacit knowledge producing knowledge claims. We have tried to demonstrate that our framework is capable to preserve organizational knowledge from being lost and most of all to create the right conditions for keeping the innovation processes.

We have choosen the model of Bianconi and Barabasi to represent the growth of the network of knowledge claims as this is the only model which allows to consider in the evolutionary process both the popularity of the externalized ideas, i.e. the number of the references made by other knowledge claims to the externalized idea, and the value or *fitness* of the externalized idea, represented by the betweeness centrality of its promoter.

Every time an individual is invited to suggest and promote a knowledge claim his/her betweeness centrality decreases favouring the augmenting of the betweeness centralities of the members of the community generated by the knowledge claim. This sort of leveling effect of the fitnesses of the externalized ideas allows to avoid the situation where a certain knowledge claim prevails over the other ones.

It is important to stress that without the particular mechanism of involving in discussions the individuals with the highest betweeness centrality, the knowledge gaps perceived by these individuals could remain internalized in a tacit form or, once externalized, could be limited to a small group of individuals strongly tied to them.

A number of issues have still to be treated.

First of all we are going to model and evaluate the effects of the introduction of discussion promoters on the overall structure [Iansiti and Levien, 2004] of the network of ideas. For example we will evaluate the robustness of the network of ideas in the presence of specific kinds of perturbations, the productivity of the network of ideas in terms of delivery of innovations and the niche creation in terms of variety, i.e. number of new ideas in a given period of time, and the overall value of the new options created. In this effort we need to take into account both local and global resources of the ecosystem. For example the fitness function should not exclusively depend on the betweeness centralities of the nodes but also on global measures of the network health as the previously introduced ones. It has been demonstrated that ecosystems governed by local and global resources can lead to the emergence of stable hubs which are a strong indicator of system robustness [Lella and Licata,2007] , and we will try to evaluate if our knowledge management model does follow this particular trend.

After this preliminary study of the knowledge model we will choice the communication channels to monitor in order to obtain a good representation of the organizational social network. Many researchers have tried to evaluate the possibility to approximate the pattern of organizational relations principally following face to face encounters, telephone communications, tele conferences, and email flows. We will review all the works regarding organizational social network analysis and we will try alternative ways to reconsruct the pattern of social ties represented by networks of k-logs. The second problem to be solved is the choice of the most suitable environment to promote the creation of knowledge. We will compare the performances of different solutions like face to face debates, forums, community blogs and wikis.

Finally we will have to define appropriate mechanisms and strategies to involve the individuals to share their knowledge and experiences, maybe suggesting them a list of possible arguments to debate with the colleagues. We will also have to define ways to invite individuals to the discussion groups. We will compare the effects of different solutions as the direct invitation of the promoter or the definiton of alerting mechanisms which for example suggest the potentially interesting discussion groups for the activities of each individual operating in the organization.

ACKNOWLEDGMENTS
This work has been partially granted by the PRIN-2005 research project "Dinamiche della Conoscenza nella Società dell'Informazione", national Coordinator Prof. Cristiano Castelfranchi.
One of authors (IL) thanks Ginestra Bianconi for her precious suggestion and encouragement.